\documentstyle[12pt]{article}

\begin{document}
\newcommand{\be}{\begin{equation}}
\newcommand{\ee}{\end{equation}}
\newcommand{\eps}{\varepsilon}

\title{Diffusion on a hypercubic
lattice with pinning potential: exact results for the
error-catastrophe problem in biological evolution.} 
\author{S. Galluccio, R. Graber and Y.-C. Zhang \\Institut de Physique 
Th\'eorique, Universit\'e de Fribourg,\\ CH -- 1700,  Switzerland}
\date{}
\maketitle

\abstract{In the theoretical biology framework one fundamental 
problem is the so-called {\it error catastrophe} 
in  Darwinian evolution models.
We reexamine  Eigen's fundamental 
equations  by mapping  them  into 
a polymer  depinning transition problem in a ``genotype'' space 
represented by a unitary hypercubic lattice $\{0, 1\}^d$. 
The exact solution of the model shows that  error 
catastrophe arises  as a direct consequence
of the equations involved and  confirms some
previous  qualitative results.  
The physically relevant consequence is that such equations
are not  adequate to properly  describe  evolution
of complex life on the Earth.} 
 
PACS numbers: 87.10.+e, 68.45.Gd
\newpage

An important question in the context of darwinian 
``natural" selection theory
is: how could complex life 
evolve and finally reach the structure
we can see nowadays by selecting the fittest
species  among  the huge number of 
different  allowed choices?
Could we explain the mechanism of self-organization
(guided evolution) to  complex life
from basic principles or is  it  necessary to consider
some other ``external"  organizing parameter?
The number $N$ of different 
realizations of  a given virus DNA chain, made of a 
very long random sequence of basic units (chemical bases), is
typically given by $N\approx 10^{1000}$.
Hence  time needed by 
random evolution
to ``explore" all possible choices before  reaching  the optimum 
sequence (i.e. complex life) is really enormous.

Here we consider a simplified model, that is evolution in a 
genotype space of dimension $d$ with a unique {\it master 
sequence}  (MS) being the favored one, i.e. the one corresponding
to individuals with highest {\it fitness}.
All other sequences are supposed to have the same lower fitness,
which, for sake of simplicity, we will take as unity.
The {\it quasi-species} \cite{E71} 
can ``diffuse'' in this genotype space
with a mutation rate per base $t$; generally assumed to be
very small.  A so-called {\it error catastrophe} arises
since increasing the chain length $d$, even though the 
master sequence $I_m$ has highest fitness, it can hardly
survive evolution. In other words, we need extremely
large fitness for $I_m$ or, equivalently, exceedingly
small mutation rate to keep the MS in a 
population.

The first investigation of this problem was 
achieved by Eigen and coworkers \cite{E71}.
 The aim of  the present work is  to solve the problem 
exactly and particular attention will be devoted to the 
conditions for occurring  of the error catastrophe.

In Eigen's model natural selection 
 is described by a simple prototype evolution equation.
  The  space of configurations, i.e. the genotype space,
 is constructed from a set ${\cal I}$ of  sequences
of uniform length comprising  $d$ monomeric units
of  which $k$ classes (chemical bases) can exist.
The number of different sequences is the 
cardinality of the set ${\cal I}$,  and obviously given by
$N=|{\cal I}|=k^d$. 
In the simplest case (the one we will consider  in the
following) $k=2$ and       
then  sequences  are made of  binary units:
$I_i=a_i \,\{ a_i=\{0,1\}, \quad \forall i=1,\cdots, d\}$.

One can then introduce 
 a continuous-time master equation
for the concentration of  individuals $x_i$ \cite{E71}.
The main result is that the target 
of selection is a species defined by the dominant, 
that is the most probable sequence $I_m$  (MS)
\cite{ES79}-\cite{S86},
which is reached  after finite time in a self-organized way, i.e. 
without any external fine tuning.
Some  heuristic  
arguments show that if the {\it excess production rate} \cite{exc} 
of the master sequence $A_m$ is too small  as compared 
with those of the mutants $A_{k\ne m}$, then
 {\it error catastrophe}  arises \cite{ECS88}:  no  convergence
to  the $I_m$ sequence takes place and  dynamics is
dominated by a random  creation and annihilation
of all possible individuals in the set ${\cal I}$.
If however  such an error catastrophe really occurred in biological
evolution, organized life couldn't  evolve on the Earth.

Our main goal in this article is  the following:
we solve exactly the evolution equations, by means
of a mapping to a polymer localization problem,
and prove that  {\it error catastrophe} 
always occurs in Eigen's model: 
the ratio $a=A_m/A_{k \ne m}$ necessary
to  self-organize the process to the master sequence
is  exponentially  big  in the  sequence length $d$. 
In other words, to localize evolution around $I_m$, Eigen's
rate equation needs an enormous selective advantage  $a$ 
(actually never realizable).            
                                    
We first define our  system and the space of configurations:
 let us consider a $d$-dimensional 
hypercubic  unitary lattice $\Omega=\{0,1\}^d$,
mimicking  a genotype space.
Each side is made of only two points  representing
binary units. Each  point of  $\Omega$ has a one-to-one 
correspondence to a sequence $I_i  \quad ( i=1,|{\cal I}|)$
since  the cardinality of ${\cal I}$ is equal to the number
of distinct points of $\Omega$ (we take $k=2$).
The discretized time version of the rate equation, 
in the polymer context, describes a depinning transition \cite{Z2}.

A polymer, directed along the time axis,
moves in a $d$-dimensional space, here $\Omega$,  
subjected to a contact energy term  with energy
gain $-u < 0$ per step of the interface
located at the wall. 
At finite temperatures $T>0$ the polymer fluctuates
in order to increase its configurational entropy but 
large fluctuations are unlikely \cite{fln91}.

The polymer is completely specified by  the Hamiltonian
\begin{equation}
{\cal H}_L\left(\{{\bf  h}\}^{(i)}\right)=
\sum_{i=1}^{L} \left(J\left|{\bf h}^{(i)}-{\bf h}^{(i)}\right|-
u\, \delta_{{\bf h}^{(i)},\vec{0}}  \right),
\end{equation}
and the partition function \cite{fln91}
\begin{equation}
{\cal Z}_L({\bf x})=\sum_{\{{\bf h}\}} \exp\left\{ 
-{\cal H}_L\left(\{{\bf h}\}^{(i)}\right) /T\right\},
\end{equation}
where ${\bf h}^{(i)}$  is the position of the polymer in $\Omega$ 
at time $i$. Overhangs are forbidden and RSOS
condition is imposed.

In order to solve  the problem we consider a transfer 
matrix approach (see also \cite{NZ95}):
\begin{equation}
{\cal Z}_{L+1}({\bf x})=
\left(1-(a-1)\delta_{{\bf x}, \vec{0}}\right)\left(
\sum_{i=1}^{d}  t{\cal Z}_{L}\left({\bf x}+{\bf e}^{(i)}\right)+
(1-dt){\cal Z}_{L}({\bf x})\right),
\end{equation}
where we have introduced the unitary vectors 
${\bf  e}^{(i)}=(0,\cdots, 1,\cdots,0)$ as those having 
a``1" bit in the $i$-th. position.
Moreover we have defined the parameters
$a=\exp(u/T)$ and $t=\exp(-J/T)$.
In this scheme the {\it mutation rate} 
$t\in [0,1]$ can be thought of as the probability
that  a given point ${\bf x}$ in $\Omega$
jumps to a well defined nearest neighbor ${\bf  x}+{\bf  e}^{(i)}$,
and  $1-dt$ is the probability that no jumps occur, that is
 $q^d=1-dt$. In the usual notation $q$ is indeed the 
probability of exact replication of  one base in the DNA chain. 

We introduce 
a dual space representation   to have  
periodic boundary conditions  in all  directions:
${\cal Z}_{L}({\bf x})=\sum_{{\bf k}=\{0,1\}^d}
(-1)^{{\bf x}\cdot {\bf k}} {\cal Z}_{L}({\bf k})$ and $
{\cal Z}_{L}({\bf k})=1/2^d\sum_{{\bf x}=\{0,1\}^d}
(-1)^{{\bf x}\cdot {\bf k}} {\cal Z}_{L}({\bf x})$. 
The summation is over the  $2^d$  possible 
binary realizations of ${\bf k}$ and ${\bf x}$.
In the dual space equation (3) takes the form
\begin{equation}
{\cal Z}_{L+1}({\bf k})=s({\bf k}){\cal Z}_{L}({\bf k})+
\frac{a+1}{2^d}\sum_{{\bf q}=\{0,1\}^d} 
s({\bf q}){\cal Z}_{L}({\bf q}),
\end{equation}
with $s({\bf q})=t\sum_{i=1}^d (-1)^{q_i}+1-dt$.
Our goal is then to solve  a $2^{d}$-dimensional eigenvalue
for  the  dual transfer matrix ${\cal M}$ acting on the
 r.h.s. of  eq. (4).
After some algebraic manipulation one can show that the
spectrum of the matrix is given by the $2^{d}$ solutions
of the following  equation:
\begin{equation}
\frac{a-1}{2^d}\sum_{{\bf k}=\{0,1\}^d} \frac{s({\bf k})}
{\eps-s({\bf k})}=1.
\end{equation}

 The search for an analytical solution 
can be simplified by recalling that in the  thermodynamic  limit
$L\rightarrow \infty$ the free energy density $f$  is dominated
by the largest eigenvalue of the spectrum, i.e. by the spectral
radius of  ${\cal M}$.

We will list below, without proof, a series of exact results;
all mathematical details will be given elsewhere \cite{gg95}.
The maximum eigenvalue of the transfer matrix ${\cal M}$
 is always {\it nondegenerate}, as a consequence of the 
Frobenius-Perron theorem, and has a corresponding positive
 right eigenvector.
We can use ${\cal M}$ to calculate, as a first approximation,
the bounds for the spectral radius $\rho({\cal M})=\eps$
by means of some theorems on positive 
matrices \cite{matr}.

We note that, strictly speaking, one  could have
a phase transition for polymer localization, i.e. 
a discontinuity in the derivatives
 of the thermodynamic potentials,  only in 
the limit $d\rightarrow \infty$.
For finite $d$ the situation is less clear. At 
any finite dimension $d$ the total number of accessible 
sites is finite and equal to $2^d$.
As a consequence our polymer never 
wanders at infinity even in the thermodynamic limit 
$t\rightarrow\infty$. 
However if the pinning strength
is not big enough the polymer is ``rough'' in the sense that
is can visit {\it all} accessible configuration space up to 
the maximum size allowed for that fixed $d$.
On the other hand, in the ``pinned'' phase, the transversal 
localization length $\ell$    within which the polymer 
is confined to the origin is independent on the linear 
size $L$ and is always finite (even at $d\rightarrow \infty$).
The two different behaviors take place at a given 
characteristic value $u_{crit}$ of the pinning potential which will
be our definition  of  criticality.
The following statements are equivalents:
in the unbounded state one has $\eps\rightarrow 1^+$,
 vanishing free energy
per unit length $f$ and constant 
components $Z(i)$ of the positive eigenvector 
associated to $\eps$. The opposite stands in the localized phase.   

Now we can turn our attention to  equation 
(5). The idea is to transform it into a  simpler
formula for $\eps$ by means of a Feynman  integral 
representation. The result is that  the maximum 
eigenvalue is given by the only real solution of the 
following implicit equation:
\begin{equation}
\frac{a}{a-1}=\eps\int_0^\infty  e^{-(\eps-1+dt)u}
\left( \cosh (ut)\right)^d\; {\rm d}u=\, 
_2F_1\left(-d; 1;\frac{\eps-1}{2t}+1;\frac{1}{2}\right).
\end{equation}
We  note that the integral  diverges iff  $\eps=1$. 
Therefore if  the  attractive potential at the origin
is omitted ($a=1$),  the maximum eigenvalue
  must be unitary, too.
Then the free energy $f$ vanishes and we attain  a delocalized 
phase, as  expected.
In the above formula $_2F_1(-a; b; c; d)$  
is the usual hypergeometric series
of negative argument $-a$ \cite{abram}.

We define $I(d; \eps, t)$ the integral in (6). The  basic
results are listed below:
In the hypothesis: $a \in (1,\infty)$;  $\eps\in (1,\infty)$;
 $t \in [0,1]$; $d\in (0,\infty)$ and $dt\in [0,1]$ then:
\begin{enumerate}
\item  $\eps(a)$ is  a convex  non-decreasing function of $a$
(strictly convex for $d$ finite). $I(d; \eps, t)$ is a 
strictly convex, decreasing  function of $d$.

\item For large $a$  the function $\eps(a; d,t)$ is linear in $a$:
$\eps\simeq (1-dt)a$,  ($a \gg1 $).
\end{enumerate}
The shape of  $\eps I(d; \eps, t)$ is showed in Fig.(2) as a 
function of $d$. Parameters $\{t,\eps\}$ are fixed in 
the physical range.

A detailed analysis of the asymptotic development 
for $I(d;\eps, t)$ at large $d$  
needs particular attention, since we should 
properly  take into account the condition $dt\le 1$.
This means that  
both the limits $d\rightarrow\infty$ and $t\rightarrow 0$
must be performed {\it simultaneously}  in the  development 
in such a way that $\alpha=dt$ be constant.
The result of the  calculation is:
\begin{equation}
\frac{a}{a-1}=\frac{\eps}{\eps-1+dt}+\frac{(dt)^2\eps}{d(\eps-1+dt)^3}
+\frac{3(dt)^4\eps}{d^2(\eps-1+dt)^5}+O\left(\frac{1}{d^3}\right).
\end{equation}
 
This implicit algebraic equation   can be solved 
for the maximum $\eps$ and  the result is compared
with the exact calculation performed by numerically finding the 
spectral radius of ${\cal M}$ for a given  set of 
parameters $\{d, t, a\}$, (see Fig.(1)).

The shape of the eigenvector corresponding to $\eps$ is
relevant from the point of view 
of the depinning transition. It is represented
 by  the sum  of its components  
$m(d;a, \eps, t)=\sum_{i}Z(i)$. One can prove
 that
\begin{equation}
m(d; a,\eps,t)=\frac{\eps}{\eps-1}\frac{a-1}{a}=
\frac{1}{\eps-1}I(d; \eps, t)^{-1}. 
\end{equation}
As a direct consequence we have that (see eq.(9) below)
$\lim_{a\rightarrow 1^+}m=2^d$ and 
 $\lim_{a\rightarrow \infty}m=1$.  
Fig.(1) shows the shape of  $m$ comparing the numerical 
result obtained by the transfer matrix and the analytical one
from the asymptotic development truncated at order 
$O(1/d^3)$. The coincidence is very good. 
Our depinning transition can be easily
studied in terms  of  $\mu=\log (m)$.
 In the unbounded state
the polymer wanders in all the accessible space of $\Omega$ 
and then $m$ reaches its maximum value,  
 while if  $a$ is very high  $\mu$ converges 
towards $0$.

If   one asks for 
the critical pinning $a_c$ necessary to localize
the polymer on the origin, we should  fix the parameters
$d$ and $t$, with the constraint $dt\le 1$, necessary 
to preserve the
probabilistic interpretation, and search for the maximum
allowed $a$ associated   with an eigenvalue $\eps$ 
``sufficiently " close  to 1. 
We specify  this  statement by considering  as  values 
close to 1 those which differ from the unity for a 
 vanishing quantity   in the limit $d\rightarrow \infty$.

This definition can be  justified, and made rigorous, by  noting
that   for $a\rightarrow 1^+$  eq.(5) is dominated
by  only one term in the sum  and one gets the result
(here $\eps$ stands for the maximum eigenvalue of the 
spectrum of ${\cal M}$) 
\begin{equation}
\frac{a}{a-1}\stackrel{a\rightarrow 1^+}{\simeq}
\frac{1}{2^d}\frac{\eps}{\eps-1}, \qquad {\rm or}\qquad
\eps\simeq 1+\frac{a-1}{1+a(2^d-1)}=1+\delta_d.
\end{equation}
Then we  can properly define $a_c={\rm sup}_{a\in (1,\infty)}
\{ a|\, \eps\le 1+\delta_d\}$.

The conclusion is that, if  $a$ is below 
$a_c$,  $\eps$ converges exponentially to $1^+$ 
in the limit $d\rightarrow \infty$.
Now we will prove the main physical result of this article, namely
that the  threshold is the critical pinning $a_c$ necessary
to localize the  polymer and that we have,  at criticality,
$\forall d$:
\begin{equation}
a_c=\frac{1+\delta_d}{1-dt} \Longleftrightarrow
d_c\simeq-\frac{\log a}{\log q},
\end{equation}
where $\delta_d$ is a function going to 0 as  $O(2^{-d})$, 
(see eq.(9)).

 The proof is rather simple if we look at the graphical 
interpretation of   eq.(6), see also Fig.(2).
 For a  given set $\{d, t, a\}$ in the physical  range, the 
nondegenerate $\eps$ is found by intersecting the curve
$\zeta=\eps I(d; a, t)$ with  the horizontal line $a/(a-1)=A=const.$.
As we showed above, $\zeta$  asymptotically converges 
to $K=\eps /(\eps-1+dt)$ for large $d$ and then to $1/dt$ in the
extreme situation $\eps \rightarrow 1^+$. 
If $a$ is too big, namely $A<K$, for a fixed $\eps$, then 
no solutions can be found since $A$ is below $\zeta$. In that
case a solution always exists but for a bigger $\eps$,
necessary to lower $K$ below $A$.
As a consequence, the critical $a_c$ following our
definition, can be 
found by asking the  maximum allowed $a$  compatible 
with a solution of  the form $\eps=1+\delta_d$.
The answer is now obvious and it is given by eq. (10).

Fig.(3) shows the critical dimension $d_c$ as a function
of  the pinning $a$ for two values of $t$.  The coincidence between
formula (10) and the  numerical results is remarkable.

Conclusion: in this Letter we have reexamined 
the evolution equations    introduced by Eigen and coworkers
in order to mimic  Darwinian natural selection in biological 
evolution.
A particle diffusing on
the   $\Omega$ space and subjected
to an attractive wall localized  at the origin
can be viewed, in the evolutionary context,
as a reproduction process  in the genotype space.
The mutation rate $t$ and the excess production rate $A_i$
for a given DNA sequence   are easily  mapped into
other physical quantities  for the polymer localization 
problem. 
Our main result is that the so-called error catastrophe
problem naturally arises as a consequence of the 
model introduced. In other words the exact solution
of Eigen equations shows that  even though a given sequence
has the highest fitness, random natural selection can
never bring evolution towards the MS  since one
 needs an extremely large fitness for the 
master sequence itself (exponentially big
in the sequence length $d$) in order to keep it surviving.
This result  suggests that Eigen's equations are
intrinsically not adequate if one is interested in the mechanisms
which explain the origin and the development 
of complex life on the Earth (characterized by individuals
of very high fitness).  
We  therefore  argue that   a  more realistic 
explanation of   this complex  phenomenology is still
lacking.  
     
 \newpage
\setlength{\baselineskip}{.6 cm}

\newpage
\section*{Figure captions}
\setlength{\baselineskip}{.8cm}

{\bf Fig. 1}\newline
Big picture: maximum eigenvalue of the transfer matrix 
${\cal M}$ plotted vs. the pinning strength $a$.
Numerical data: full line; analytical result (up
to order $O(1/d^3)$): circles. The dashed lines are the bounds
for $\eps$  obtained from the transfer matrix.
Small picture: $\log[m(a)]$ vs. $a$ . In all cases  
$d=100$,  $t=.003$.
Numerical data: full line. Analytical result: circles.
\vskip .7cm
\noindent
{\bf Fig. 2}\newline
Shape of $I(d;\eps,t)$ and of $a/(a-1)$ vs. 
$d$ (see text). The dashed line gives the asymptotic limit of $I$ for
large dimensions $d$ at $\eps \rightarrow 1^+$.
\vskip .7cm
\noindent
{\bf Fig. 3}\newline
Critical dimension $d_c$ vs. pinning strength $a$ for
two distinct values of $t$. Lower curve: $t=10^{-2}$;
upper curve: $t=10^{-3}$.
Full lines represent  the function
$d_c=t^{-1}(1-1/a)$ (see text). 
Circles and squares: numerical data from the transfer matrix. 
\end{document}